\documentclass[format=acmsmall,manuscript=true,review=false,nonacm,screen=true]{acmart}

\setcopyright{acmcopyright}
\copyrightyear{2022}

\newcommand{\solidfmm}{\texttt{solidfmm}}

\begin{document}

\title[\solidfmm: An optimised library of solid harmonics for use in FMMs]%
{\solidfmm: A highly optimised library of operations on the solid
harmonics for use in fast multipole methods}

\author{Matthias Kirchhart}
\email{kirchhart@acom.rwth-aachen.de}
\orcid{0000-0001-7018-6099}
\affiliation%
{
  \institution{Applied and Computational Mathematics, RWTH Aachen}
  \streetaddress{Schinkelstra{\ss}e~2}
  \city{52062~Aachen}
  \country{Germany}
}

\begin{abstract}
We present \solidfmm, a highly optimised C++ library for the solid harmonics as
they are needed in fast multipole methods. The library provides efficient,
vectorised implementations of the translation operations M2M, M2L, and L2L, and
is available as free software. While asymptotically of complexity $O(P^3)$, for all practically relevant expansion orders $P$, the translation operators 
display an empirical complexity of $O(P^2)$, outperforming the na\"{\i}ve
implementation by orders of magnitude.
\end{abstract}



\maketitle

\section{Introduction}
\subsection{Summary}
\solidfmm\ is a highly optimised C++ library for operations on the solid
harmonics as they are needed in fast multipole methods.

\begin{itemize}
\item Supports both single and double precision.
\item Fully vectorised on x86 CPUs with support for AVX or AVX-512.
\item Hand-written assembly implementations for maximal performance of critical
      parts.
\item Supports mixed order translations: you can do M2M, M2L, L2L with
      differing input and output orders. This is important for adaptive, variable order fast multipole codes.
\item Translations support arbitrary shift vectors.
\item No dependence on external libraries except for the C++ standard library.
\item Thread safe and exception safe.
\item Lean. The library interface is very small and easy to use. We used
      encapsulation using handles to avoid dependence of user code on
      implementation details.
\item Free software. Licenced under the GNU General Public License as published
      by the Free Software Foundation, either version 3, or, at your option,
      any later version.
\end{itemize}

\solidfmm\ is available from the author's institution at
\url{https://rwth-aachen.sciebo.de/s/YIJFvSERVBiOkbc}, or from GitHub.

\subsection{Motivation and Background}
The fast multipole method (FMM)~\cite{greengard1987} has been called one of the
top ten algorithms of the 20th century~\cite{cipra2000}. It allows for
approximate solutions of $N$-body problems, namely the evaluation of:
\begin{equation}\label{eqn:nbody}
u(x_i) = \sum_{j=1}^{N}G(x_i,x_j)q_j\qquad i=1,\dots,N,
\end{equation}
where $G$ is the so-called kernel function, $N$ is the number of bodies with
locations $x_i\in\mathbb{R}^3$ and associated masses or charges
$q_i\in\mathbb{R}$, $i=1,\dotsc,N$. It is straightforward to see that a direct
evaluation of this sum at all particle locations requires $O(N^2)$ arithmetic
operations, making this approach infeasible for large numbers of bodies $N$. In
contrast, the FMM computes approximations of \eqref{eqn:nbody} in only $O(N)$
operations, resulting in dramatic speedups, where the accuracy can be
controlled by the user.

This is achieved through the use of series expansions of the underlying kernel
function $G$. These expansions are truncated at a user-defined order $P$, where
increasing  $P$ results in more accurate, but also more expensive computations:
the hidden constant in the $O(N)$-complexity depends on $P$ and the type of
expansion used.  One of the most important kernel functions is the fundamental
solution of the Poisson equation, i.\,e., up to a factor of $(4\pi)^{-1}$, the
function $G(x,y)\coloneqq |x-y|^{-1}$, $x,y\in\mathbb{R}^3$. It can be used for
both gravitational as well as electrical potentials. The solid harmonics have
been specifically derived with this particular kernel in mind and yield the
most efficient expansions available. Expansions in their terms require only 
$O(P^2)$ coefficients, compared to $O(P^3)$ for Chebyshev or Cartesian Taylor
expansions at the same level of accuracy. Depending on the application, common
values for $P$ lie between $P\approx 3$ for fast, low-fidelity computations and
$P\approx 30$ to satisfy the highest accuracy demands. Thus, by modern
standards, an individual expansion is `small', and millions of such expansions
can be stored in the memory of even cheap hardware.

There are two different kinds of expansions in FMMs: mutipole and local
expansions, also called M- and L-expansions, respectively. In the so-called
M2L-translation,\footnote{Strictly speaking, the term `translation' is a 
misnomer as it also involves a conversion. Yet, the name got stuck and is now
commonly used in the context of FMMs.} an M-expansion around one point $A$
is converted into an L-expansion around another point $B$. This operation is
without doubt the most crucial element of FMMs, for large $N$ it is required
millions of times and most of the computational effort lies in this stage. 
The direct implementation of the M2L-translation has complexity $O(P^4)$.
This number is not at all that small anymore when $P\approx 30$ and
millions of such operations are necessary. For this reason, several
so-called `fast translation' schemes have been devised reducing this
complexity to $O(P^3)$ or even $O\bigl(P^2(\log_2 P)^2\bigr)$.

On the one hand, many of the commonly available FMM software projects focus on
the difficult  task of creating an implementation that efficiently scales to
highly parallel super-computers. Very elaborate techniques are used to minimise
communication, handle load balancing, and to additionally make use of GPUs when
available. On the other hand, the M2L-translation in these software packages is
often implemented using simple, nested loops without further optimisation.
ExaFMM~\cite{yokota2013b}, for example, uses the straight
$O(P^4)$ implementation\footnote{See \url{%
https://github.com/exafmm/exafmm/blob/master/kernels/laplace.h}}
and additionally uses the trigonometric functions to evaluate the solid
harmonics in spherical coordinates---even though they can more easily be
computed in Cartesian coordinates using only square roots and
elementary arithmetic~\cite{wang1996}. Such implementations thus leave an
opportunity for improvement at the local, CPU level. For this reason we believe
that even sophisticated codes like ExaFMM could directly benefit from an
optimised implementation of the M2L translation.

The main reason for this situation is probably that implementing fast,
efficiently vectorised translation schemes for the solid harmonics is quite
challenging. To our knowledge there simply is no implementation of fast
translation schemes that is free (as both in beer and in liberty) and
vectorised. \solidfmm\ aims to change this situation. It provides a
fully vectorised  implementation of an algorithm sketched by
Dehnen~\cite{dehnen2014} with highly optimised routines for current x86
architectures. Inspired by the BLIS~\cite{vanzee2015}, the performance critical
parts are reduced to a handful of isolated and small `microkernels' written in
assembly language and compiler intrinsics, while the main part  of the library
is written in platform-independent standard C++. This extensible design allows
us to add support for further CPU architectures in the future. In addition
to the M2L, the library also features accelerated implementations of the other
translations in FMM: M2M and L2L. The library is intended to be used the
context of the aforementioned highly-parallel FMM frameworks, such that they
immedialetely benefit from its optimisations.

In this article we will first summarise the basics of the M2L translation.
The M2M and L2L translations are very similar in nature, and while also
implemented in \solidfmm, omitted here for brevity.  We then continue
with a very brief review of various acceleration techniques, before describing
Dehnen's approach in greater detail. We then describe how \solidfmm\
implements a slightly modified version of his algorithm and finish with
benchmarks comparing \solidfmm\ to the simple, na\"{\i}ve implementation.

\section{The M2L Translation}
In this section we give a review of the M2L translation. After describing its
plain, canonical formulation we give a short survey of different acceleration
techniques, before describing the ideas behind the current, rotation-based
approach in greater detail.

\subsection{Canonical Formulation}
M- and L-expansions of order $P$, developed around respectively
$x_A,x_B\in\mathbb{R}^3$ are functions of the following shape:
\begin{equation}\label{eqn:MLexpansions}
\sum_{n=0}^{P-1}\sum_{m=-n}^{n} M_n^m \overline{S_n^m(x-x_A)},\quad
\sum_{n=0}^{P-1}\sum_{m=-n}^{n} L_n^m \overline{R_n^m(x-x_B)}.
\end{equation}
Here, $R_n^m$ and $S_n^m$ are complex-valued functions, respectively
called the \emph{regular} and \emph{singular harmonics}, as defined in the
appendix. The coefficients $M_n^m\in\mathbb{C}$ are also called multipoles,
whereas the $L_n^m\in\mathbb{C}$ are simply called local coefficients.

For $C\in\lbrace L,M,R,S\rbrace$ one always has $C_n^{-m} =
(-1)^m\overline{C_n^m}$. This ensures that the sums in~\eqref{eqn:MLexpansions}
always take real values; additionally one only needs to store
the coefficients with $m\geq 0$. Further space could be saved by noting that
for $m=0$ the imaginary parts $\Im(C_n^0)$ are always vanishing. However, the
memory savings from this are only marginal and come at the cost of more
complicated memory layouts. For this reason, in this work, we will also store
$\Im(C_n^0)=0$.

\begin{figure}
\includegraphics[scale=0.8]{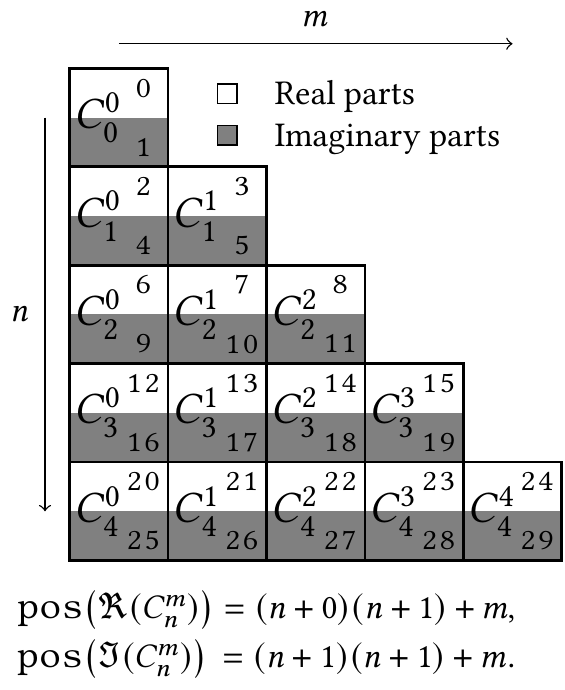}
\Description{A triangular array that illustrates how coefficients of
both M- and L-expansions are stored in memory.}
\caption{\label{fig:triangle}The coefficients $C_n^m\in\mathbb{C}$ of both
multipole and local expansions can be stored using $P(P+1)$ real numbers. In
this example we have $P=5$.  They can be arranged in a triangular pattern
according to their indices $n=0,\dotsc,4$ and $m=0,\dotsc,n$, where the values
for negative $m$ are given implicitly by $C_n^{m}=(-1)^m\overline{C_n^{-m}}$.
The numbers in the upper and lower right corners of the boxes denote the array
position (\texttt{pos}) in computer memory. Throughout this work, we will refer
to such an arrangement of numbers as a \emph{solid}.}
\end{figure}

Thus, an M- or L-expansion of order $P$ can be stored using $P(P+1)/2$
complex, or  $P(P+1)$ real numbers in a `triangular shape' as illustrated
in \autoref{fig:triangle}. Lacking a better name, we will call such a
triangular arrangement of coefficients a \emph{solid}. We thus have defined
solids and harmonics.\footnote{For the rest of this work, we slightly deviate
from the standard mathematical nomenclature, where the functions $R_n^m$ and
$S_n^m$ themselves are called solid harmonics. For us, the functions are called
\emph{harmonics}; a set of their values at a specific point or a set of
coefficients is called \emph{solid}.}

Given a multipole expansion with coefficients $M_n^m$ and expansion centre
$x_A$, the corresponding local expansion at centre $x_B$ and with coefficients
$L_n^m$ can be computed using the M2L-translation. It comes in two flavours:
\begin{equation}\label{eqn:m2lkernels}
L_n^m = (-1)^n \sum_{k=0}^{P-n-1}\sum_{l=-k}^{k} \overline{M_k^l}
S_{n+k}^{m+l}(x_B-x_A)
\qquad\text{and}\qquad
L_n^m = (-1)^n \sum_{k=0}^{P-1}\sum_{l=-k}^{k} \overline{M_k^l}
S_{n+k}^{m+l}(x_B-x_A),
\end{equation}
respectively called the `single height' and `double height' kernels. These
expressions only differ in the number of terms: for the double height
kernel we need to compute $S$ up to order $2P-1$, while for the single height
kernel order $P$ suffices. The single height kernel is cheaper, the
double height kernel is more accurate. In its na\"{\i}ve implementation,
the double height kernel is about six times more expensive to evaluate than
the single height version~\cite[Footnote~5]{coulaud2008}. This should be kept
in mind when comparing implementations.

\solidfmm\ implements the double height kernel, as it leads to more
regular memory access patterns: every entry in $L$ linearly depends on every
entry in $M$. Thus, M2L can be interpreted as a linear mapping, i.\,e., a
fully populated matrix, whose $O(P^4)$ entries $S_{n+k}^{m+l}(x_B-x_A)$ depend
on the shift vector $r_{AB}\coloneqq x_B-x_A$.

\subsection{A Brief Survey of Fast M2L Schemes}
Various approaches have been proposed to accelerate the M2L translation,
a survey of the most common ones is given by Coulaud, Fortin, and Roman~%
\cite{coulaud2008}. Some of these approaches only work when there only is a
finite set of shift vector directions $r_{AB}\coloneqq x_B - x_A$, which is
the case in certain implementations of the FMM. These approaches include:
\begin{itemize}
\item BLAS-based approaches~\cite{coulaud2008}. These do not break the $O(P^4)$
complexity.  Instead, the finite set of translation matrices is stored
explicitly, performance is achieved by using highly optimised BLAS routines to
apply these matrices to many solids in parallel.
\item Plane-wave expansion methods~\cite{greengard1997}. These only approximate
the M2L translation and need careful tuning to match the accuracy of the
original expansions for every given $P$. 
\end{itemize}
These schemes are unsuitable for use in \solidfmm, which aims to be
flexible and permit arbitrary shifts $r_{AB}$ and orders~$P$.

Other, more general approaches include:
\begin{itemize}
\item Diagonalising the translation matrix through the use of fast Fourier
transforms (FFT)~\cite{elliott1996}. These approaches achieve the aforementioned
complexity of $O\bigl(P^2(\log_2 P)^2\bigr)$. However, practice has shown that
they suffer from numerical instabilities for $P>16$. 
\item Rotation-based approaches~\cite{white1996}. These perform a change of
coordinates, after which the translation can be carried out in $O(P^3)$ time.
Finally, the result is changed back to the original coordinate system. Such
changes of coordinates can also be done in $O(P^3)$ time, resulting in the same
overall complexity of $O(P^3)$.
\end{itemize}
There are approaches to repair the instabilities of the FFT method, but they
all come at an extra cost. Additionally, we note that $(\log_2 P)^2\geq P$ for
$4\leq P\leq 16$, so for the stable choices of $P$, whether the algorithmic
complexity is favourable depends on the hidden constants and even more on the
particular implementation. For these reasons \solidfmm\ implements a
rotation-based approach, which will be described in greater detail below.

\subsection{Rotation-based Approaches}
Rotation-based accelerations were introduced by White and Head-Gordon~%
\cite{white1996}. The basic idea is as follows. Assume the shift-vector $r_{AB}$
was aligned with the z-axis, i.\,e., assume we had $r_{AB}=(0,0,R)^\top$. In
this case, most of the entries in the M2L-kernels~\eqref{eqn:m2lkernels} vanish,
because:
\begin{equation}
S_{n+k}^{m+l}(0,0,R) =
\begin{cases}
0 & \text{if $m+l\neq 0$}, \\
\frac{(n+k)!}{R^{n+k+1}} & \text{if $m+l=0$}.
\end{cases}
\end{equation}
We now add a minor modification, by additionally assuming that $R=1$.
Then~\eqref{eqn:m2lkernels} reduces to:
\begin{equation}\label{eqn:zm2lkernels}
L_n^m = (-1)^{n+m} \sum_{k=m}^{P-n-1}M_k^m(n+k)!
\qquad\text{and}\qquad
L_n^m = (-1)^{n+m} \sum_{k=m}^{P-1}M_k^m(n+k)!
\end{equation}
In other words, for $r_{AB}=(0,0,1)^\top$, M2L is an operation that
acts \emph{column-wise} on a solid, resulting in $O(P^3)$ complexity.
Thus, for the double height kernel, the general structure of the algorithm is
as follows. Given $\nu\in\mathbb{N}$ solids $M[0], M[1],\dotsc, M[\nu-1]$ of
order $P$ and associated shift vectors $r[0], r[1],\dotsc, r[\nu-1]$, do the
following:
\begin{enumerate}
\item Perform a change of coordinates on each $M[0],\dotsc,M[\nu-1]$, such
that the shift is along $(0,0,1)^\top$.
\item For each column $m=0,\dotsc,P-1$, perform the following matrix--matrix
product:
\begin{equation}\label{eqn:zm2l-matrix}
\begin{pmatrix}
2m!      & (2m+1)! & (2m+2)!  & \cdots & (P+m-1)! \\
(2m+1)!  & (2m+2)! & (2m+3)!  & \cdots & (P+m)!   \\
\vdots   & \vdots  & \vdots   & \ddots & \vdots   \\
(P+m-1)! & (P+m)!  & (P+m+1)! & \cdots & (2P-2)!
\end{pmatrix}
\begin{pmatrix}
M_m^m[0]     & M_m^m[1]     & \cdots & M_m^m[\nu-1]      \\
M_{m+1}^m[0] & M_{m+1}^m[1] & \cdots & M_{m+1}^m[\nu-1]  \\
\vdots       & \vdots       & \ddots & \vdots            \\
M_{P-1}^m[0] & M_{P-1}^m[1] & \cdots & M_{P-1}^m[\nu-1]
\end{pmatrix}.
\end{equation}
Note that real and imaginary parts do not couple! This matrix--matrix
product can be carried out separately for real and imaginary parts. Also
note that we apply the same matrix of faculties to all solids.
\item Apply the signs $(-1)^{n+m}$ to the results.
\item Reverse the change of coordinates back to the original.
\end{enumerate}

Here we see why we introduced the additional assumption $R=1$: otherwise
each solid would need to be multiplied with a different matrix. Now the
the matrix is constant and identical for all solids $M[0],\dotsc,M[\nu-1]$. The
matrix--matrix product~\eqref{eqn:zm2l-matrix} can be highly optimised using
blocking techniques from the BLIS~\cite{vanzee2015} and can be carried out
separately for real and imaginary parts. However, to avoid wasting precious
cache space, the matrix of faculties should not be stored explictly. It
suffices to store the the values $0!,1!,\dotsc,(2P-2)!$ in a vector and write a
specialised routine to carry out the product~\eqref{eqn:zm2l-matrix} using this
vector.

Previous authors did not carry out the scaling to $R=1$, and left the lengths of
the shift vectors  $r_{AB}$ unchanged. In this case the necessary change of
coordinates corresponds to a rotation, hence the name `rotation-based'. We
will discuss different, but equivalent methods to achieve these rotations
in the following subsection.

\subsection{Rotating and Scaling Coordinate Systems}
While M2L along the z-axis is an operation that acts column-wise on a solid, a
change of coordinates acts row-wise. Scaling is trivial: when transforming a
vector as $r\mapsto sr$ for some $s>0$, the M- und L-coefficients
change as follows:
\begin{equation}\label{eqn:scaling}
M_n^m\mapsto s^{n+1}M_n^m,\qquad L_n^m\mapsto \frac{L_n^m}{s^n}.
\end{equation}
The cost of this operation is negligible. Letting $s = |r|^{-1}$ the
shift vector gets unit length, and it remains to perform rotations.

\subsubsection{General Rotations}
A general rotation takes the following form:
\begin{equation}\label{eqn:generalrotation}
M_n^m\mapsto\sum_{l=-n}^{n}D^{l,m}_nM_n^l,\quad
L_n^m\mapsto\sum_{l=-n}^{n}D^{m,l}_nL_n^l,
\end{equation}
where the $D_n^{m,l}$ are the so-called Wigner matrices. These matrices depend
on the Euler angles of the rotation in a highly non-trivial manner and its
entries are very expensive to evaluate. For this, we refer to the original paper
of White and Head-Gordon~\cite{white1996}. If, however, only a finite set of
shift directions $r_{AB}$ is permitted, these matrices can be precomputed.
Again, this makes this approach unsuitable for use in \solidfmm, which
aims to allow arbitrary shift vectors $r_{AB}$.

\subsubsection{Factorisation of the Wigner Matrix}
The idea of factorising the Wigner matrices $D_n^{m,l}$ into simpler matrices
goes back to Wigner himself. It has long been known that rotations around the
z-axis are trivial. For any angle $\alpha$, consider the transformation:
\begin{equation}
(x,y,z)^\top \mapsto
\bigl (x\cos(\alpha) + y\sin(\alpha),
      -x\sin(\alpha) + y\cos(\alpha), z\bigr)^\top.
\end{equation}
Under such rotations, the M- and L-coefficients transform as follows:
\begin{equation}
M_n^m \mapsto M_n^m e^{im\alpha},\qquad L_n^m \mapsto L_n^m e^{im\alpha}.
\end{equation}

Wigner suggested precomputing the matrices $D_n^{m,l}$ for rotations by
90\textdegree\ around the $y$-axis. This way, a general rotation can be
carried out by combining three trivial rotations around the z-axis with two
rotations by 90\textdegree\ around the $y$-axis. In other words, we can make
use of precomputed matrices $D_n^{m,l}$ for 90\textdegree, independent of the
actual Euler angles of the rotation. Just like for the z-translation in
\eqref{eqn:zm2l-matrix}, the same matrices $D_n^{m,l}$ are applied to all
solids. The apparent drawback of this approach is that now two $O(P^3)$
operations of the shape~\eqref{eqn:generalrotation} are necessary instead of
just one. 

\subsubsection{Dehnen's Factorisation}
Dehnen~\cite{dehnen2014} picked up this idea in the context of M2L, and
considered swapping the $x$- and $z$-axes instead. The corresponding
matrices are denoted $B_n^{m,l}$, resulting in the transformations
\begin{equation}\label{eqn:swaptransform}
M_n^m\mapsto\sum_{l=-n}^n B_n^{l,m}M_n^l,\quad
L_n^m\mapsto\sum_{l=-n}^n B_n^{m,l}L_n^l.
\end{equation}
Swapping two axes turns a right-handed coordinate system into a left-handed
one. However, because this operation is carried out twice, one obtains properly
oriented results in the end. This approach has several benefits:
\begin{enumerate}
\item The matrices $B_n^{m,l}$ are involutory, i.\,e., they are their own
inverses. Afterall, swapping x and z, and then swapping them again will get
you back to where you started. There thus is no need to separately store
the inverses.
\item All entries $B_n^{m,l}$ are real-valued. Thus, real and imaginary parts
decouple in \eqref{eqn:swaptransform} and can be computed separately, cutting
the operation count in half. 
\item The matrices for the real and imaginary transformations are each only
50\% populated. This reduces the cost of their application again by half and
will be described in more detail later.
\end{enumerate}
Thus Dehnen's approach requires swapping the x- and z-axes twice, but each
of these swaps only cost $O(\tfrac{1}{2}P^3)$ operations, making the
approach competitive to precomputing the Wigner matrices $D_n^{m,l}$ for
a set of fixed directions, but without the associated loss of generality.

\subsubsection{Summary}
In total, combining Dehnen's approach with rescaling reduces the complexity 
of the M2L translation from $O(P^4)$ to $O(P^3)$ by factorising the operation
into:
\begin{itemize}
\item $O(P^3)$ operations, namely the swapping of axes and translation
along $(0,0,1)^\top$. These operations are `expensive', but independent of the
shift vectors. These operations can thus be vectorised and highly optimised by
writing them as matrix--matrix products.
\item $O(P^2)$ operations, namely scaling and rotations around the z-axis. These
do depend on the shift vectors, but their cost is negligible. Additionally,
vectorisation of these operations is even easier than vectorisation of the
matrix--matrix products.
\end{itemize}
This approach is thus ideally suited for efficient implementations modern CPUs.
Dehnen mentions that he implemented a vectorised version of this scheme, but
to our knowledge his code is not freely available.

\section{Implementation}

\subsection{Precomputation}
The precomputation stage consists of computing the matrices required for
swapping co-ordinate axes and storing them in a suitable memory layout. The
computation of the faculties $0!,1!,\dotsc,(2P-2)!$ is trivial, and they can
be stored in a simple vector. The matrices $B_n^{m,l}$ were introduced by
Dehnen and we repeat his recurrence relations in the appendix. After these have
been computed the corresponding matrices for the real and imaginary parts of
solids can be computed. Using the facts that $M_{n}^{m} =
(-1)^m\overline{M_n^{-m}}$ and $B_n^{m,l}\in\mathbb{R}$, the 
mappings~\eqref{eqn:swaptransform} become:
\begin{equation}
\Re(M_n^m)\mapsto B_n^{0,m}\Re(M_n^0) +
\sum_{l=1}^n\bigl(B_n^{l,m} + (-1)^l B_n^{-l,m}\bigr)\Re(M_n^l),
\quad
\Im(M_n^m)\mapsto\sum_{l=1}^n\bigl(B_n^{l,m}-(-1)^l B_n^{-l,m}\bigr)\Im(M_n^l),
\end{equation}
and analogous for $L_n^m$. We thus define the \emph{swap matrices} $F_n^{m,l}$
and $G_n^{m,l}$, $m,l=0,\dotsc,n$ as follows:
\begin{equation}
F_n^{m,l}\coloneqq
\begin{cases}
B_n^{0,m}                     & \text{if $l=0$,} \\
B_n^{l,m} + (-1)^l B_n^{-l,m} & \text{else,}
\end{cases}
\qquad
G_n^{m,l}\coloneqq
\begin{cases}
0                             & \text{if $l=0$ or $m=0$,} \\
B_n^{l,m} - (-1)^l B_n^{-l,m} & \text{else.}
\end{cases}
\end{equation}
With this, the mappings~\eqref{eqn:swaptransform} turn into:
\begin{equation}\label{eqn:FGmappings}
\begin{alignedat}{2}
\Re(M_n^m)&\mapsto\sum_{l=0}^n F_n^{m,l}\Re(M_n^l), &\qquad
\Im(M_n^m)&\mapsto\sum_{l=0}^n G_n^{m,l}\Im(M_n^l), \\
\Re(L_n^m)&\mapsto\sum_{l=0}^n F_n^{l,m}\Re(L_n^l), &\qquad
\Im(L_n^m)&\mapsto\sum_{l=0}^n G_n^{l,m}\Im(L_n^l).
\end{alignedat}
\end{equation}

\begin{figure}
\centering
\includegraphics[scale=0.5]{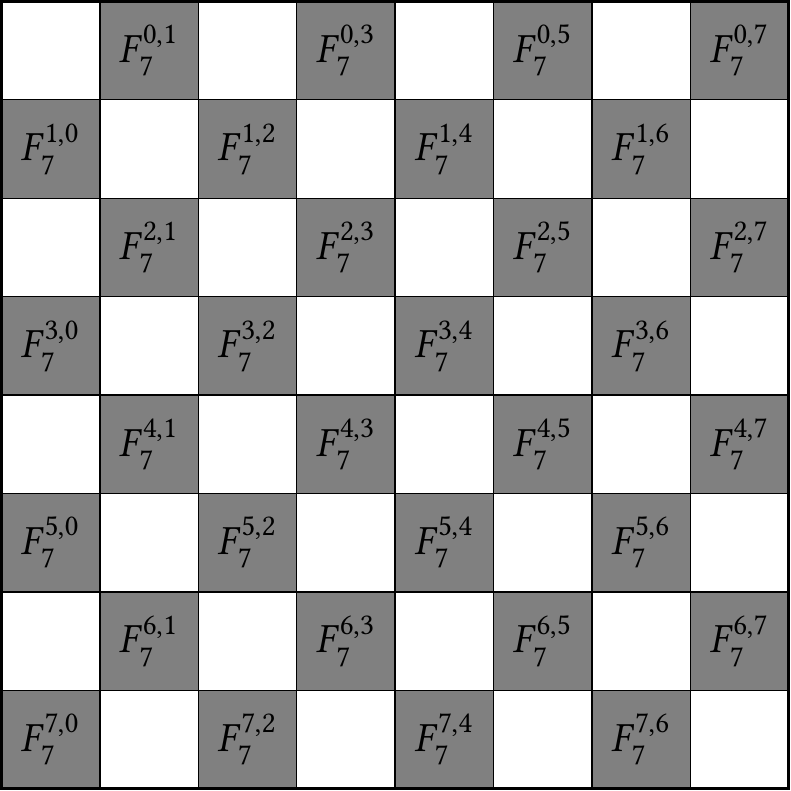}
\qquad
\includegraphics[scale=0.5]{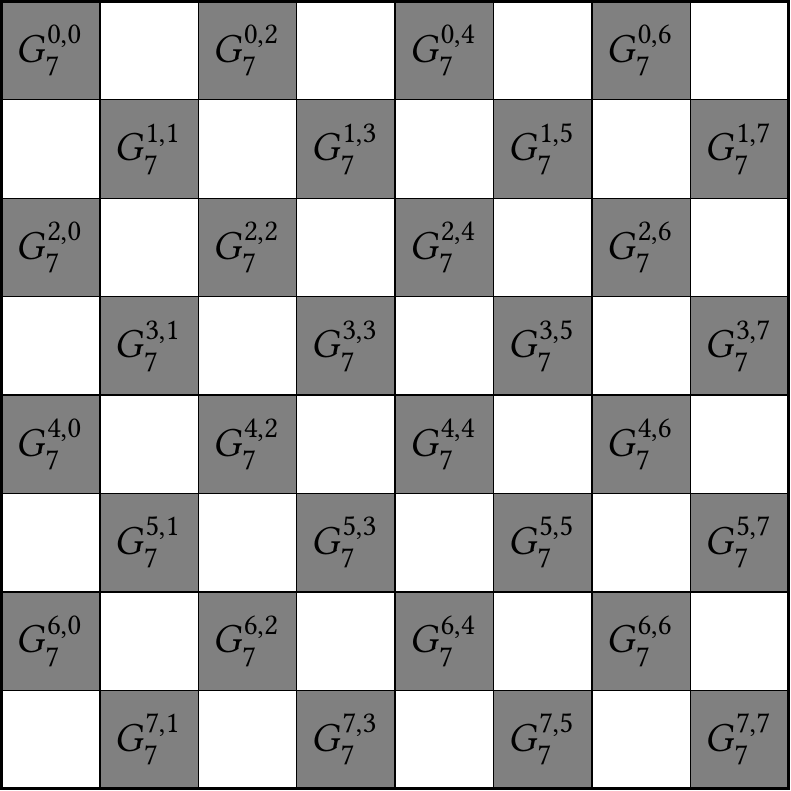}
\caption{\label{fig:chequerboard}The swap matrices $F_n^{m,l}$ (left) and 
$G_n^{m,l}$ (right), here illustrated for $n=7$, exhibit a chequerboard
pattern: every other entry is zero (white). For $F_7$ the pattern begins with
zero, $G_7$ shows the opposite pattern. For even values of $n$ the situation is
reversed. Only the non-zero values should be stored to save memory and optimise
the use of cache space. (More precisely, the first row and column of $G_n$ are
always entirely zero and the pattern only shows in the remaining part of the
matrix. The overhead of considering these additional zeros, however, is
marginal, and keeping them leads to simpler and more uniform code.)}
\end{figure}

When storing the matrices $F_n^{m,l}$ and $G_n^{m,l}$, as well as when
implementing the mappings~\eqref{eqn:FGmappings}, it is important to notice
that only 50\% of their entries are non-zero. In particular the matrices exhibit
a chequerboard pattern, as shown in \autoref{fig:chequerboard}.

\subsection{Blocking and Packing}
The computationally most expensive parts of the M2L translation consist of
dense matrix--matrix products. We can thus build on the experience of the
BLIS~\cite{vanzee2015} developers and apply blocking techniques. In particular,
when forming a matrix--matrix product $C=AB$, the matrix $C$ is sub-divided
into blocks of size $\mu\times\nu$, where the values of $\mu$ and $\nu$ are
machine dependent. These blocks are then computed using the SIMD instructions
of the processor, which are encapsulated in a so-called microkernel. A
microkernel itself thus is machine specific, but it offers a generic, platform
independent interface. The blocking scheme can thus be written using generic,
platform independent code which then calls the machine specific microkernel to
perform the actual computation. If necessary, $A$ and $C$ are padded by
additional zero rows such that the total number of rows is a multiple of $\mu$.

On many x86 systems supporting the AVX instruction set the optimal block size
for double precision is $\mu = 6$ and $\nu=8$, where $B$ and $C$ are stored in
row-major order.  On a single core of such machines, we can thus process
$\nu=8$ solids in double precision in parallel.\footnote{An AVX register can
only hold four double precision values. However, most Intel microarchitectures
since Haswell can simultaneously execute arithmetic instructions on two such
registers in a single clock cycle, giving eight values in total.}
On systems supporting AVX-512 we choose $\mu=14$ and $\nu=16$. Unlike the BLIS,
we do not need to introduce an entire hierarchy of blocks that matches the cache
hierarchy, as for realistic orders $P$ the swap matrices $F_n$ and $G_n$ are
comparatively small and completely fit into the L2 and L3 caches of current
CPUs.

\begin{figure}
\centering
\includegraphics[scale=0.8]{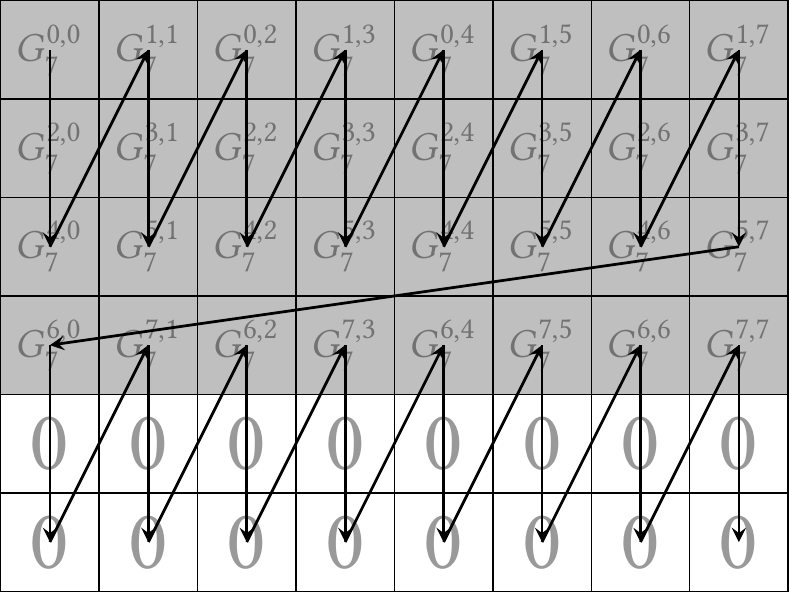}
\caption{\label{fig:compressedmatrix}On a system with $\mu=6$, the matrix $G_7$
would be stored in memory as shown above. First, $G_7$ would be padded by with
additional zero rows, such that the total number of rows is a multiple of
$\mu$. In this case four rows of zeros are padded, resulting in
$G_{7,\textrm{padded}}\in\mathbb{R}^{12\times8}$. Then the first
$\mu$ rows are stored in column-major order, omitting the zeros from the
chequerboard pattern. The next $\mu$ rows follow, again omitting zeros according
to the chequerboard pattern. When computing a matrix--matrix product with
$G_7$, the entries $G_7^{m,l}$ are accessed in the exact same order. We thus
benefit from hardware prefetching and decreasing L1 cache misses. This
technique is known as `packing'~\cite[Section~3.3]{laffblis}. This requires
that $G_n$ and $F_n$ and their respective transposes $G_n^\top$ and $F_n^\top$
are each packed and stored separately.}
\end{figure}

While for the product \eqref{eqn:zm2l-matrix} the matrix $A$ of the generic
product $C=AB$ can be compressed into a single vector of faculties, for the swap
matrices $F_n$ and $G_n$ it is stored in the so-called packed format as
illustrated in \autoref{fig:compressedmatrix}.

\subsection{Buffers and Rearranging}
Axis swap operations act row-wise on a solid and do not mingle real and
imaginary parts. Thus, given $\nu$ solids and row $n$, we need to compute the
products:
\begin{equation}
F_n \cdot
\begin{pmatrix}
\Re(M_n^0[0]) & \Re(M_n^0[1]) & \cdots & \Re(M_n^0[\nu-1]) \\
\Re(M_n^1[0]) & \Re(M_n^1[1]) & \cdots & \Re(M_n^1[\nu-1]) \\
\vdots        & \vdots        & \ddots & \vdots            \\
\Re(M_n^n[0]) & \Re(M_n^n[1]) & \cdots & \Re(M_n^n[\nu-1])
\end{pmatrix},
\quad
G_n \cdot
\begin{pmatrix}
\Im(M_n^0[0]) & \Im(M_n^0[1]) & \cdots & \Im(M_n^0[\nu-1]) \\
\Im(M_n^1[0]) & \Im(M_n^1[1]) & \cdots & \Im(M_n^1[\nu-1]) \\
\vdots        & \vdots        & \ddots & \vdots            \\
\Im(M_n^n[0]) & \Im(M_n^n[1]) & \cdots & \Im(M_n^n[\nu-1])
\end{pmatrix}.
\end{equation}

In practice these matrices of real and imaginary parts will need to be stored
in row-major order and padded with zero rows, such that the total number of rows
is a multiple of $\mu$. We thus need two buffers for holding these matrices,
which are then filled with data from solids that may lie scattered in arbitrary
positions in memory. We will call these the swap buffers.

On the other hand, the z-translation acts column-wise on a solid. In order to
carry it out efficiently, we need to store the matrices:
\begin{equation}
\begin{pmatrix}
\Re(M^m_m[0])     & \Re(M^m_m[1])     & \cdots & \Re(M^m_m    [\nu-1]) \\
\Re(M^m_{m+1}[0]) & \Re(M^m_{m+1}[1]) & \cdots & \Re(M^m_{m+1}[\nu-1]) \\
\vdots            & \vdots            & \ddots & \vdots                \\
\Re(M^m_{P-1}[0]) & \Re(M^m_{P-1}[1]) & \cdots & \Re(M^m_{P-1}[\nu-1])
\end{pmatrix},
\quad
\begin{pmatrix}
\Im(M^m_m[0])     & \Im(M^m_m[1])     & \cdots & \Im(M^m_m    [\nu-1]) \\
\Im(M^m_{m+1}[0]) & \Im(M^m_{m+1}[1]) & \cdots & \Im(M^m_{m+1}[\nu-1]) \\
\vdots            & \vdots            & \ddots & \vdots                \\
\Im(M^m_{P-1}[0]) & \Im(M^m_{P-1}[1]) & \cdots & \Im(M^m_{P-1}[\nu-1])
\end{pmatrix},
\end{equation}
for each $m=0,1,\dotsc,P-1$; padded and in row-major order. We will call these
the translation buffers. In order to achieve maximum performance, it is thus
necessary to copy data from the swap buffers into the translation buffers and
vice versa. This is unfortunate, but there seems no way around this rearranging
of data in memory.

Here the benefit of the row-major format becomes obvious. One row from a
swap buffer corresponds to exactly one row in the corresponding translation
buffer. We can increase performance by choosing $\nu$ such that one row
corresponds to a fixed number of entire cache lines. For x86 systems with AVX
support, this achieved by the choice $\nu=8$ for double precision: one row
of these matrices then exactly corresponds to 64 bytes, the size of a cache
line. We can thus always copy entire cache lines at once instead of proceeding
element-wise in a scalar, non-vectorised fashion.
 
\subsection{Summary: The Entire Algorithm}
We now have gathered all the main ingredients of the algorithm implemented in
\solidfmm, and are in a position to summarise it here. Thus, let there
be $\nu$ solids $M[0],\dotsc,M[\nu-1]$ and shift vectors $r_{AB}[0],\dotsc,
r_{AB}[\nu-1]$ be given. We then proceed as follows, where for brevity we will
omit the indices $[0],[1],\dotsc,[\nu-1]$ from now on:
\begin{enumerate}
\item For $n=0,\dotsc,P$, and for each $\nu$ compute the following quantities:
$|r_{AB}|^n$, $|r_{AB}|^{-n}$, $e^{in\alpha}$, $e^{in\beta}$. The exponential
terms are powers of
$e^{i\alpha} =\cos\alpha + i\sin\alpha$, $e^{i\beta}=\cos\beta + i\sin\beta$,
where:
\begin{equation}
\begin{alignedat}{2}
\cos\alpha&=\frac{y}{\sqrt{x^2+y^2}},     &\qquad
\cos\beta &=\frac{z}{|r_{AB}|}, \\
\sin\alpha&=\frac{x}{\sqrt{x^2+y^2}},     &\qquad
\sin\beta &=-\frac{\sqrt{x^2+y^2}}{|r_{AB}|}.
\end{alignedat}
\end{equation}
Thus, no trigonometric functions are necessary for this computation. Special
care needs to be taken for $\alpha$ when $x=y=0$. In this case: $\cos\alpha=1$
and $\sin\alpha=0$.
\item For each row $n=0,\dotsc,P-1$:
\begin{enumerate}
\item Copy row $n$ from all of the $\nu$ solids into the swap buffers.
\item Scale and rotate around $z$ axis with angle $\alpha$:\ 
\begin{math}
M_n^m\mapsto\frac{M_n^m}{|r_{AB}|^{n+1}}e^{im\alpha},\ m=0,1,\dotsc,n.
\end{math}
\item Swap x- and z-axes:
\begin{equation}\label{eqn:Mswapxz}
\Re(M_n^m)\mapsto \sum_{l=0}^{n}F_n^{m,l}\Re(M_n^l),\quad
\Im(M_n^m)\mapsto \sum_{l=0}^{n}G_n^{m,l}\Im(M_n^l),\qquad m=0,1,\dotsc,n.
\end{equation}
Only every other entry in the matrices $F_n^{m,l}$, $G_n^{m,l}$, $l=0,\dotsc,n$
is non-zero.
\item Rotate around $z$ axis by angle $\beta$:\ 
\begin{math}
M_n^m\mapsto M_n^m e^{im\beta},\ m=0,1,\dotsc,n.
\end{math}
(No scaling)
\item Swap $x$- and $z$-axes again, according to~\eqref{eqn:Mswapxz}.
\item Copy the result into the translation buffers.
\end{enumerate}
\item For each column $m=0,\dotsc,P-1$: compute the matrix--matrix product
\eqref{eqn:zm2l-matrix}, apply the signs $(-1)^{n+m}$.
\item For each row $n=0,\dotsc,P-1$:
\begin{enumerate}
\item Gather row $n$ for each of the $\nu$ solids from the translation buffers
and store them into the swap buffer.
\item Swap the $x$- and $z$-axes:
\begin{equation}\label{eqn:Lswapxz}
\Re(L_n^m)\mapsto\sum_{l=0}^{n}F_n^{l,m}\Re(L_n^l),\quad
\Im(L_n^m)\mapsto\sum_{l=0}^{n}G_n^{l,m}\Im(L_n^l),\qquad m=0,1,\dotsc,n.
\end{equation}
Here the transposed matrices $F_n^{l,m}$, $G_n^{l,m}$ are used.
\item Rotate around $z$ axis by the negative angle $-\beta$:\ 
\begin{math}
L_n^m\mapsto L_n^m e^{-im\beta},\ m=0,1,\dotsc,n.
\end{math} (No scaling.)
\item Swap x- and z-axes again, according to equation~\eqref{eqn:Lswapxz}.
\item Rotate around $z$ axis by $-\alpha$ and undo the scaling:\ 
\begin{math}
L_n^m\mapsto \frac{L_n^m}{|r_{AB}|^n}e^{-im\alpha},\ m=0,1,\dotsc,n.
\end{math}
Note that the denominator has power $n$, not $n+1$.
\item Store the result at the desired output location.
\end{enumerate}
\end{enumerate}

\section{Performance Measurements}
At the moment, \solidfmm\ comes in version 1.2 with generic microkernels and
optimised kernels for AVX and AVX-512. We consider the timings for the M2L kernel only, as it is the most crucial of the FMM's translations. The
benchmarks can be carried out using the executables \texttt{benchmark\_m2l} and
\texttt{benchmark\_m2l\_naive} for the  accelerated and na\"{\i}ve
implementations, respectively. These programmes carry out a large number of M2L
translations for each order $P=1,\dotsc,50$ and measure the time needed to
complete the task. Afterwards this amount of time is divided by the number of
translations. The na\"{\i}ve implementation uses simple, four-fold nested loops
to implement the double height kernel~\eqref{eqn:m2lkernels}.

\begin{figure}
\includegraphics{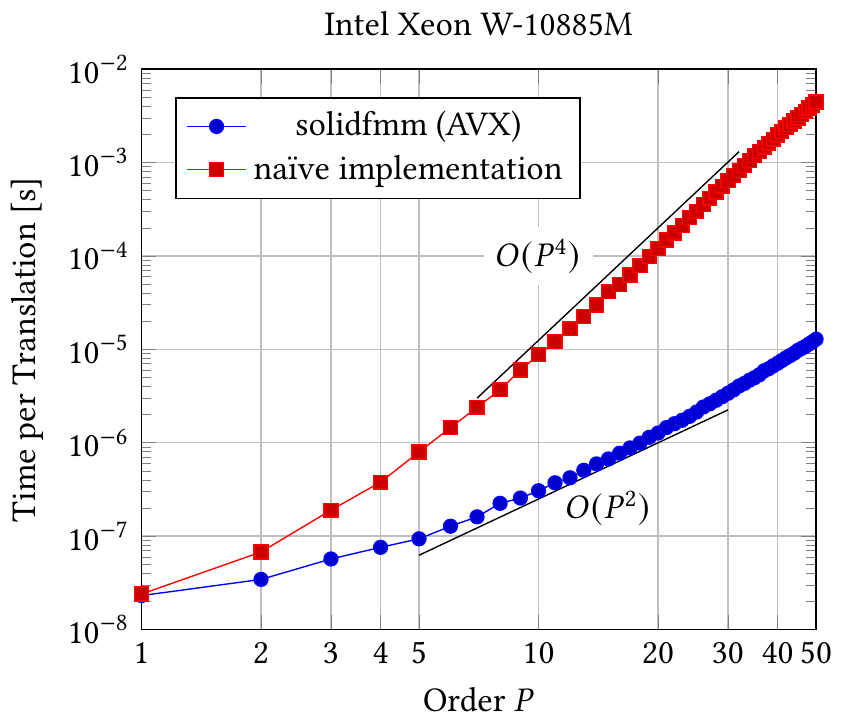}\vspace{1cm}
\includegraphics{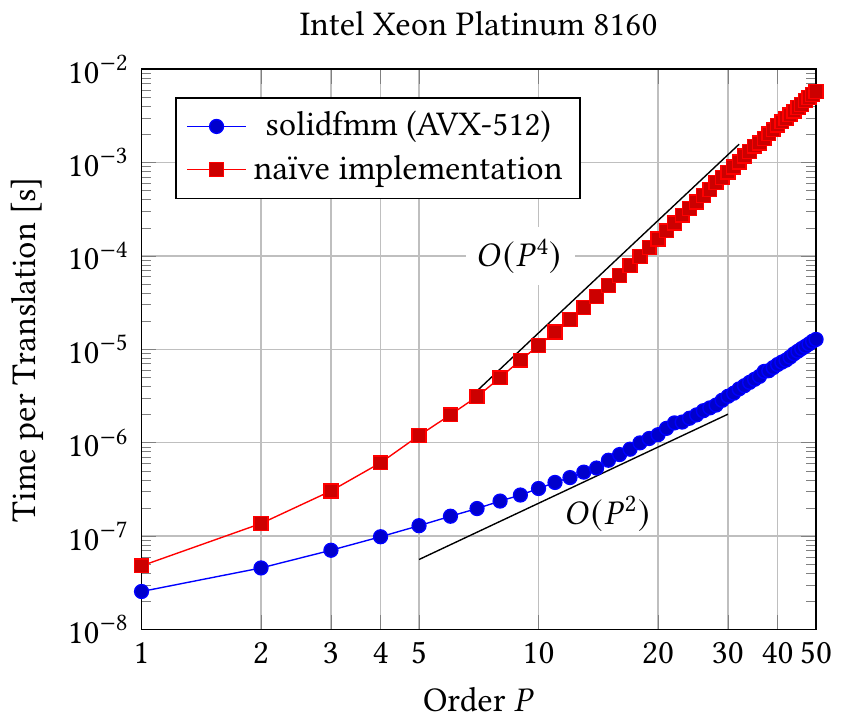}
\caption{\label{fig:m2ltimings}Timings for performing a single M2L translation 
of order $P$, for a na\"{\i}ve, loop-based implementation and \solidfmm\ using
double precision arithmetic, performed on a single core of an Intel Xeon
W-10885M (top) and an Intel Xeon Platinum 8160 (bottom). The timings were
obtained by carrying out a large number of M2L translations, the total time was
then divided by this number. \solidfmm\ is vectorised using respectively
AVX and AVX-512 with $\nu=8$ and $\nu=16$, the na\"{\i}ve implementation is scalar
and uses a simple, four-fold nested loop to implement the double height
kernel~\eqref{eqn:m2lkernels}.\ One clearly sees the $O(P^4)$ complexity of the
na\"{\i}ve code. \solidfmm, on the other hand, shows an empirical complexity of
$O(P^2)$ or better for the practically relevant orders, clearly outperforming the
asymptotic bound of $O(P^3)$. For the AVX-512 version, the effect of the blocking
scheme ($\mu=14$) becomes apparent at orders $P=14$ and $P=28$, where the
slope starts to slightly increase.}
\end{figure}

We carried out experiments on a single core of an Intel Xeon W-10885M, which
supports AVX, and an Intel Xeon Platinum 8160, supporting AVX-512. The results
are illustrated in \autoref{fig:m2ltimings}. The na\"{\i}ve implementation shows
the expected $O(P^4)$ complexity. \solidfmm, on the other hand, performs faster
than expected: we observe $O(P^2)$ or better for all relevant orders, in
contrast to the asymptotic complexity of $O(P^3)$. We believe that his is due to
the following effects. The microkernels only achieve a fraction of their peak
performance when applied to small problems. Thus, the growing complexity is
compensated by higher computational efficiency. At the largest orders we then
finally begin to see an increased slope. For the AVX-512 version one can clearly
observe the effects of the blocking scheme with $\mu=14$: at order $P=14$ the
slope increases to about $O(P^2)$, another increase is visible at $P=28$.

Already at $P=10$ we observe a 33-fold speed up for the Xeon Platinum 8160,
while for the W-10885M it is 28. These numbers keep on increasing, at $P=20$
the respective speed ups are 124 and 94. We therefore believe that \solidfmm\
is particularly interesting for high accuracy computations.

\begin{acks}
This research was carried out under funding of the German Research Foundation
(DFG), project `Vortex Methods for Incompressible Flows', grant number
432219818. Without their support, this project would not have been possible.

I also received funding from the German National High Performance Computing
(NHR) organisation. 

I would also like to acknowledge the work of Simon Paepenmöller, one of my
student workers. He helped in the development of preliminary software designs,
and the tracking of many nasty sign bugs. This work is completely new, but
draws from conclusions from these initial attempts and would not have been
possible without them.
\end{acks}

\appendix
\section*{Appendix}
Here, for completeness, we repeat the recurrence relations as given by
Dehnen~\cite{dehnen2014}.

\subsection*{Regular and Singular Harmonics}
Let $r=(x,y,z)^\top$ be given. Starting with $S_0^0(r)=|r|^{-1}$ and
$R_0^0(r)=1$, one continues with the diagonal:
\begin{equation}
S_n^n = (2n-1)\frac{x+iy}{|r|^2}S_{n-1}^{n-1},\qquad
R_n^n =       \frac{x+iy}{2n}R_{n-1}^{n-1},
\end{equation}
and then obtains the remaining entries via:
\begin{equation}
\begin{split}
|r|^2 S_n^m    &= (2n-1)zS_{n-1}^m-\bigl((n-1)^2-m^2\bigr)S_{n-2}^m,\\
(n^2-m^2)R_n^m &= (2n-1)zR_{n-1}^m - |r|^2R_{n-2}^m.
\end{split}
\end{equation}

\subsection*{The Matrices $B_n^{m,l}$}
Starting with $B_0^{0,0}=1$, one has:
\begin{equation}
2B_{n+1}^{m,l} = B_n^{m,l-1} - B_n^{m,l+1}, \qquad
2B_{n+1}^{m\pm1,l} = B_n^{m,l-1} \pm 2 B_n^{m,l} + B_n^{m,l+1},
\end{equation}
where it is implicitly assumed that $B_n^{m,l}=0$, whenever $|l|>n$.

\bibliographystyle{ACM-Reference-Format}
\bibliography{literature.bib}
\end{document}